\documentclass[aps,pre,reprint,superscriptaddress]{revtex4-2}

\usepackage[T1]{fontenc}
\usepackage{amsmath,amssymb}
\usepackage{graphicx}
\usepackage{bm}
\usepackage{xcolor}
\usepackage{url}

\begin{document}

\title{Normal-Field Evolution and the Breakdown of Classical Tearing in Current
Sheets}

\author{Grzegorz Kowal}
\email{grzegorz.kowal@usp.br}
\affiliation{Escola de Artes, Ci\^encias e Humanidades, University of S\~ao
Paulo, S\~ao Paulo, SP, Brazil}

\author{Diego A. Falceta-Gon\c{c}alves}
\email{dfalcetal@usp.br}
\affiliation{Escola de Artes, Ci\^encias e Humanidades, University of S\~ao
Paulo, S\~ao Paulo, SP, Brazil}

\date{\today}
\begin{abstract}
Classical resistive tearing theory assumes a prescribed current-sheet
equilibrium, but a finite normal magnetic field can evolve the background on
the tearing timescale. For a transversely pressure-balanced Harris profile of
initial half-thickness $a$ and asymptotic reconnecting-field magnitude $B_0$,
threaded by uniform $B_n$ ($\xi\equiv B_n/B_0$), we derive an exact evolution in
ideal magnetohydrodynamics: counter-propagating Alfv\'enic fronts broaden the
neutral layer and generate a sheet-parallel flow with velocity gradients on
its two flanks. The same normal field enters the linearized perturbation
equations directly through ideal transport, alongside advection and
flow-curvature terms from the evolving velocity profile. Inner-layer estimates
recover
$\xi_{\rm crit}\sim S^{-3/4}$ for the breakdown of the resistive balance at
the classical fastest-growing scale, where $S$ is the Lundquist number based
on $a$. Numerical frozen-profile calculations show that growing modes can
persist after central resistive dominance is lost; eigenfunction comparisons
at fixed wavenumber and predominantly Doppler-shifted propagation support a
normal-field-modified continuation of classical tearing. The normal field
supplies no net perturbation energy. Energy extraction from the
reconnecting-field gradient dominates at small and moderate widths and weakens
as the sheet broadens. The velocity-gradient contribution becomes comparable
only in broad, weakened modes, where it partly offsets the loss without
reversing the decline in growth. For the representative case $S=10^6$, magnetic
Prandtl number
$\mathrm{Pr}_m=0$, and $\xi=10^{-4}$, the maximum growth rate falls by a factor
of about ten from its initial value when the neutral-layer half-width $w$
reaches $w/a\simeq1.59$, where the growth and broadening times become equal.
Only about $0.18$ additional upper-envelope e-folds are estimated between this
crossing and $w/a=2$, where the growth rate is about 25 times smaller than its
initial value and the instantaneous spectrum remains unstable. These results
support dynamical suppression of further linear amplification over the
sampled evolution, while nonlinear island or plasmoid formation remains
unresolved. Current-sheet disruption criteria must therefore account for both
direct modification of the inner balance and normal-field-driven background
evolution.
\end{abstract}

\maketitle

\section{Introduction}

The resistive tearing instability provides one of the canonical routes by which
magnetic fields reconnect in thin current sheets, forming magnetic islands or
plasmoids \cite{1963PhFl....6..459F,2007PhPl...14j0703L,2009PhPl...16k2102B,
2016PhPl...23j0702C}. Its classical description assumes a prescribed
equilibrium magnetic profile. The perturbation obeys ideal MHD in the
outer region, where the linear equations are singular at the resonant surface,
and resistivity resolves that singularity within a narrow inner layer.
Matching the two solutions introduces the tearing parameter $\Delta'$, the jump
in the logarithmic derivative of the ideal outer magnetic perturbation across
the resonant surface, which measures the drive available to the inner layer.
This construction underlies much of the modern discussion of current-sheet
disruption, including tearing-mediated turbulence and plasmoid-mediated
reconnection
\cite{2017PhRvL.118x5101L,2017MNRAS.468.4862M,2022PhRvX..12d1027G}.

Island-like or flux-rope-like structures in simulations
\cite{2007PhPl...14j0703L,2009PhPl...16k2102B,2016PhPl...23j0702C,
2017PhRvL.118x5101L,2022PhRvX..12d1027G}
and in observations of reconnecting current sheets
\cite{2020NatCo..11.3964W,2022NatCo..13..640Y,2022FrASS...9.9491B} are often
interpreted as evidence of tearing-mediated disruption. Morphology alone,
however, does not establish that a structure formed through the classical
mechanism, which requires a sufficiently long-lived background, an appropriate
resonant structure, and a resistivity-dominated inner layer that can be matched
to ideal outer solutions
\cite{1963PhFl....6..459F,1976SvJPP...2..533C,2014ApJ...780L..19P}. In evolving
current sheets these assumptions must be checked rather than inferred from
plasmoid-like features. Disruption criteria for plasmoid reconnection,
ideal tearing, and tearing-mediated turbulence accordingly compare the tearing
growth time with the lifetime of a sheet or eddy
\cite{2014ApJ...780L..19P,2017PhRvL.118x5101L,2022PhRvX..12d1027G}, but they
usually hold the background fixed while the perturbation grows and so
can neglect background evolution driven by a weak normal
field.

Previous studies incorporate shear or background evolution in
physically distinct ways. Chen and Morrison and Ofman et al. analyzed tearing
about prescribed, time-independent magnetic and equilibrium shear-flow profiles
\cite{1990PhFlB...2..495C,1991PhFlB...3.1364O}. Forming-sheet theories instead
follow tearing while an imposed driver increases the sheet aspect ratio:
Uzdensky and Loureiro treated disruption during a generic driven evolution and
an explicit Chapman--Kendall example, Comisso et al. selected the mode
completing its linear phase first during a prescribed thinning history, and
Tolman, Loureiro, and Uzdensky evolved a narrowing and strengthening sheet under
an incompressible stagnation-point flow
\cite{2016PhRvL.116j5003U,2016PhPl...23j0702C,2018JPlPh..84a9015T}; in the
turbulence setting, Mallet et al. compared tearing-mediated disruption of
dynamically aligned sheet-like structures with their cascade time
\cite{2017MNRAS.468.4862M}. The present mechanism has no external thinning
driver: the finite $B_n$ both generates the analytically linked
reconnecting-field and streamwise-flow profiles and remains explicitly in the
perturbation operator. The sheet broadens rather than thins, and the resulting
ideal terms can overtake resistive induction, so background evolution can change
the inner balance itself rather than only make the classical tearing rate time
dependent.

Three-dimensional MHD turbulence provides one setting in which current
sheets arise self-consistently as intermittent structures
\cite{2013ApJ...771..124Z,2022SciA....8N7627D}. Solenoidality requires the
magnetic component normal to an idealized sheet to be continuous across it but
not to vanish, so turbulent sheets need not be exact tangential discontinuities.
How common finite local $B_n$ is, and how its distribution depends on sheet
geometry, lifetime, and plasma parameters, remain separate questions. The
present work is therefore conditional: it determines the local dynamics when
$B_n$ is finite and approximately uniform, while a statistical assessment is
deferred to a companion numerical study.

Finite normal fields have been reported or constructed in several
current-sheet settings. A Cluster survey found $B_z$ at magnetotail
current-sheet centers to be mostly positive and typically
$1$--$10\,\mathrm{nT}$ \cite{2011JGRA..116.9218R}; a global hybrid-Vlasov
simulation produced a thin magnetotail current sheet with a small normal
component before ion-mediated tearing and kink instabilities developed
\cite{Zaitsev2025}; and one-dimensional kinetic current-sheet solutions with a
small normal field have been constructed from particle-orbit invariants and
pressure anisotropy \cite{2023PhPl...30k2903E}. These results motivate
considering local current-sheet regions threaded by finite normal fields, but
their geometries, closures, and dynamics differ from the present model, and none
shows that the uniform-$B_n$ ideal-MHD evolution derived here has been observed
or reproduced globally.

A normal field is already known to modify tearing in a prescribed
background. Somov and Verneta showed that motion across it produces an opposing
Lorentz force that changes the resistive inner-layer balance in a quasi-steady
current layer
\cite{1988SoPh..117...89S,1989SoPh..120...93S,1993SSRv...65..253S}, and in their
thickness-based Lundquist-number ordering this correction becomes stabilizing
when $\xi=B_n/B_0$ is comparable to $S^{-3/4}$. We use that scale only as an
order-of-magnitude reference, since its numerical coefficient and detailed
ordering need not carry over to the evolving-sheet problem.

Here we test whether the prescribed-background ordering survives when
the same $B_n$ both evolves the sheet and enters the perturbation operator.
Section~II derives the exact ideal-MHD evolution of a transversely
pressure-balanced Harris profile in the presence of a normal magnetic field.
Section~III formulates the instantaneous tearing eigenproblem for the
resulting profiles. Section~IV derives the
expected modifications relative to classical tearing, including the far-field
spatial structure and inner-balance transition scales. Section~V presents
representative numerical spectra and diagnostics of the inner-balance
transition. Section~VI examines the physical character of the growing mode
using its eigenfunction structure, comoving frequency, and
perturbation-energy budget. Section~VII discusses the physical implications
and summarizes the conclusions.

\section{Harris Current Sheet in the Presence of a Normal Field}
\label{harris}

First, we pose an initial-value problem for a one-dimensional Harris
current sheet of initial half-thickness $a$
\cite{1962NCim...23..115H}. The reconnecting
field is initialized as $B_x(t=0,z)=B_0\tanh(z/a)$, with no initial in-plane
flow and a uniform normal component $B_z=\mathrm{const}$. A guide-field or
thermal-pressure profile may be used to maintain transverse
total-pressure balance; the tearing calculation below depends only on the closed
$(U_x,B_x)$ response. The relevant one-dimensional MHD equations are
\begin{align}
\frac{\partial U_x}{\partial t}
&=B_z\frac{\partial B_x}{\partial z}
+\nu\frac{\partial^2 U_x}{\partial z^2},
\label{eq:ux}\\
\frac{\partial U_y}{\partial t}
&=B_z\frac{\partial B_y}{\partial z}
+\nu\frac{\partial^2 U_y}{\partial z^2},
\label{eq:uy}\\
\frac{\partial U_z}{\partial t}
&=-\frac{\partial}{\partial z}
\left[p+\frac{1}{2}\left(B_x^2+B_y^2\right)\right],
\label{eq:uz}\\
\frac{\partial B_x}{\partial t}
&=B_z\frac{\partial U_x}{\partial z}
+\eta\frac{\partial^2 B_x}{\partial z^2},
\label{eq:bx}\\
\frac{\partial B_y}{\partial t}
&=B_z\frac{\partial U_y}{\partial z}
+\eta\frac{\partial^2 B_y}{\partial z^2},
\label{eq:by}\\
\frac{\partial B_z}{\partial t}&=0.
\label{eq:bz}
\end{align}
with magnetic fields expressed in Alfv\'en-speed units, $V_A\equiv B_0$.
Equation~(\ref{eq:bz}) shows that the normal component is constant; below we
denote it by $B_n\equiv B_z$ and keep $\xi\equiv B_n/B_0$ throughout, so that
the Alfv\'enic crossing time along the normal field is $a/B_n=a/(\xi V_A)$.
At $t=0$ the Harris profile and the chosen pressure or guide-field
profile balance forces in the transverse $z$ direction, but the configuration is
not stationary when $B_n\ne0$: the term $B_n\partial B_x/\partial z$ in
Eq.~(\ref{eq:ux}) is a tangential magnetic-tension force that accelerates the
initially static plasma. Equations~(\ref{eq:ux}) and (\ref{eq:bx}) therefore
describe an ideal Alfv\'enic relaxation that broadens the reversing field and
generates a sheet-parallel flow with velocity gradients concentrated in two
flank regions, rather than a sequence of stationary equilibria.
Incompressibility and the choice $U_z=0$ make Eq.~(\ref{eq:uz}) a
transverse total-pressure constraint,
$p+(B_x^2+B_y^2)/2=\mathrm{const}$, so changes in $B_x$ may be compensated by
thermal pressure, guide-field pressure, or both without changing this
closed $(U_x,B_x)$ response. For a guide field such as
$B_y(t=0,z)=B_0\operatorname{sech}(z/a)+B_g$, the $(U_y,B_y)$ subsystem
undergoes the same ideal propagation along $B_n$, given by Eqs.~(\ref{eq:Bx})
and (\ref{eq:Ux}) with $\operatorname{sech}$ replacing $\tanh$ and $B_g$ added
to $B_y$, while Eq.~(\ref{eq:uz}) fixes the thermal-pressure adjustment that
keeps the total pressure uniform; this closure does not feed back on
Eqs.~(\ref{eq:ux}) and (\ref{eq:bx}) in the two-dimensional incompressible
problem.

The resulting solution is a local model for a pre-existing or
dynamically formed current-sheet region threaded by a small normal field. Its
use requires the region to remain approximately planar over the scales and times
considered, $B_n$ to be nearly uniform across it, the sheet thickness to remain
within the fluid-MHD regime, the broadening time $a/B_n$ to be competitive with
the tearing growth time, and boundaries or global geometry not to suppress the
counter-propagating Alfv\'enic response. It is not a global model of a
finite-length, compressible, or kinetic current sheet, and we do not claim that
its exact analytical profiles have been observed, reproduced in global
simulations, or realized in a laboratory plasma. Subject to these conditions,
it isolates whether local normal-field-driven background evolution invalidates
the fixed-profile tearing ordering.

Finite viscosity and resistivity are analytically tractable---for $\nu=\eta$
the Elsasser variables $U_x\pm B_x$ obey advection-diffusion equations---but,
because the corresponding Reynolds numbers are large in typical astrophysical
sheets, these corrections are small on the Alfv\'enic broadening time
considered here. We therefore take $\nu=\eta=0$ to obtain the closed-form
evolving background used here; finite viscosity can still alter the
tearing-layer scalings in the eigenvalue problem below.
Taking the time derivative of Eq.~(\ref{eq:bx}) and using Eq.~(\ref{eq:ux}) then
gives
\begin{equation}
\frac{\partial^2 B_x}{\partial t^2}
=B_n^2\frac{\partial^2 B_x}{\partial z^2}.
\end{equation}
For the initial Harris profile, the solution is
\begin{equation}
\label{eq:Bx}
\begin{aligned}
    B_x(t,z)
    &=\frac{B_0}{2}
    \left[
        \tanh\left(\frac{z+w(t)}{a}\right)
        +\tanh\left(\frac{z-w(t)}{a}\right)
    \right],\\
    w(t)&\equiv B_n t=\xi V_A t .
\end{aligned}
\end{equation}
The solution describes a current sheet that develops a neutral layer of
vanishing $B_x$ around the midplane, of half-width $w(t)$, or equivalently
$w/a=\xi V_A t/a$ in units of the initial sheet half-thickness. We take $\xi>0$
without loss of generality; reversing $B_n$ reverses the streamwise
flow and the signs of its flank gradients, as well as the propagation
direction, but leaves the broadened $B_x$ profile unchanged. The
streamwise flow $U_x$ follows from Eq.~(\ref{eq:ux}) by integrating
the $z$ derivative of $B_x$ over time,
{
\begin{equation}
\label{eq:Ux}
    U_x(t,z)
    =\frac{B_0}{2}
    \left[
        \tanh\left(\frac{z+w(t)}{a}\right)
        -\tanh\left(\frac{z-w(t)}{a}\right)
    \right],
\end{equation}
}%
which vanishes at $t=0$. The streamwise velocity reaches its maximum
at $z=0$, where $U_x'(t,0)=0$, so velocity gradients develop on the two flanks
of the neutral layer:
\begin{equation}
    U_x(t,0)
    =B_0\tanh\left(\frac{w(t)}{a}\right),
\end{equation}
so the central-flow amplitude initially grows linearly at a rate
proportional to $\xi V_A/a$ and then saturates at the Alfv\'en speed after
$t_{\rm sat}\simeq a/(\xi V_A)$. The stability calculation below therefore uses
only the self-consistent reconnecting-field and streamwise-flow
profiles in Eqs.~(\ref{eq:Bx}) and (\ref{eq:Ux}).

\section{Tearing Instability Analysis of the Evolving Sheet}
\label{stability}

We now freeze this exact evolving profile at a given $w$ and examine its
instantaneous tearing spectrum. Let $\sigma$ denote the complex eigenvalue of
the linearized problem and $\gamma\equiv\operatorname{Re}\sigma$ the growth
rate. The quasi-steady approximation is valid when the tearing growth time is
shorter than the profile-evolution time, i.e. when $\gamma w/(\xi V_A)\gg1$.
Since $w(t)=\xi V_A t$, the local broadening time is $t_w=w/(\xi V_A)$, so
equality of the instantaneous growth and broadening timescales gives
the local comparison $\gamma\gtrsim\xi V_A/w$, which in the normalized units
used below becomes $\gamma\gtrsim\xi/w$ for $w>0$. Modes below this
threshold remain valid instantaneous eigenmodes but fail this local survival
test, which alone does not determine their accumulated amplification. A
stricter frozen-profile ordering would require a larger margin, e.g.
$\gamma\gg\xi/w$, and a global survival test would use the accumulated
amplification along the evolving sequence. With $\widehat w\equiv w/a$ and
$\tau_A\equiv a/V_A$, let $k$ denote the sheet-parallel wavenumber. The
corresponding number of e-foldings estimated from the instantaneous
eigenvalues is
{\bfseries
\begin{align}
    N(k,\widehat w)
    &\equiv \int_0^{t(\widehat w)}
    \gamma[k,w(t')]\,dt' \notag\\
    &=\frac{1}{\xi}\int_0^{\widehat w}
    \gamma(k,\widehat w')\tau_A\,d\widehat w',
    \label{eq:fixed_k_amplification}\\
    N_{\max}(\widehat w)
    &\equiv\frac{1}{\xi}\int_0^{\widehat w}
    \gamma_{\max}(\widehat w')\tau_A\,d\widehat w'.
    \label{eq:max_amplification}
\end{align}
}
where $\gamma_{\max}(\widehat w)\equiv\max_k\gamma(k,\widehat w)$. The
first expression follows one fixed Fourier component; since
$\gamma(k,\widehat w)\leq\gamma_{\max}(\widehat w)$ at every width,
$N(k,\widehat w)\leq N_{\max}(\widehat w)$, so the second is an upper envelope
rather than the temporal history of one mode when the maximizing wavenumber
varies with $w$. When the local ordering is not large, the eigenproblem is best
interpreted as an instantaneous diagnostic of the branches supported by the
evolving profile rather than as a strictly time-asymptotic modal calculation.
Let
\begin{align}
    F(z) &=
    \frac{1}{2}
    \left[
        \tanh\left(\frac{z+w}{a}\right)
        +\tanh\left(\frac{z-w}{a}\right)
    \right], \\
    G(z) &=
    \frac{1}{2}
    \left[
        \tanh\left(\frac{z+w}{a}\right)
        -\tanh\left(\frac{z-w}{a}\right)
    \right],
\end{align}
so that $F$ and $G$ are the dimensionless reconnecting field and
streamwise flow, respectively. Hereafter,
$\boldsymbol{u}=(u_x,u_y,u_z)$ and $\boldsymbol{b}=(b_x,b_y,b_z)$ denote the
velocity and magnetic-field perturbations, normalized by $V_A$ and $B_0$,
respectively. For perturbations proportional to $\exp(\sigma t+ikx)$,
incompressibility gives $b_x=i b_z'/k$ and $u_x=i u_z'/k$. With
$\mathcal{D}\equiv d^2/dz^2-k^2$, the independent linearized equations are
\begin{align}
    \sigma \mathcal{D}u_z
    &= ik\left(F\mathcal{D}b_z-F''b_z\right) \notag\\
    &\quad
    -ik\left(G\mathcal{D}u_z-G''u_z\right) \notag\\
    &\quad
    +\xi(\mathcal{D}b_z)'
    + \mathrm{Pr}_m S^{-1}\mathcal{D}^2u_z,
    \label{eq:lin_u}\\
    \sigma b_z
    &= ik\left(Fu_z-Gb_z\right)
    + \xi u_z'
    + S^{-1}\mathcal{D}b_z.
    \label{eq:lin_b}
\end{align}

Here $S=aV_A/\eta$ is the Lundquist number based on the initial current-sheet
half-thickness, as is customary in tearing-mode theory,
$\mathrm{Pr}_m\equiv\nu/\eta$ is the magnetic Prandtl number, so that the
dimensionless viscous coefficient is $\mathrm{Pr}_m S^{-1}$, and primes denote
derivatives with respect to $z$. The profiles $F$ and $G$ are the analytical
solutions for $B_x/B_0$ and $U_x/V_A$ evaluated at the evolutionary stage
$w=\xi V_A t$. In the
eigenvalue problem, lengths entering derivatives are normalized by $a$,
wavenumbers by $a^{-1}$, and eigenvalues by $\tau_A^{-1}$; explicit factors of
$a$ are retained in profile definitions and scaling estimates for clarity.

\section{Expected Modifications to the Classical Tearing Structure}
\label{subsec:tearing_structure}

It is useful to separate how the additional terms in
Eqs.~(\ref{eq:lin_u})--(\ref{eq:lin_b}) modify the classical tearing-mode
structure. In the Furth--Killeen--Rosenbluth (FKR) constant-$\psi$
Harris-sheet problem, the ideal outer solution is singular at the resonant
surface and the resistive term in Eq.~(\ref{eq:lin_b}) resolves this singularity
inside a narrow inner layer. Here the induction equation contains the additional
ideal contribution $\xi u_z'$, while the equilibrium flow contributes
through advection, $-ikG b_z$, in Eq.~(\ref{eq:lin_b}) and through the
vorticity-advection and flow-curvature terms, $-ikG(u_z''-k^2u_z)$
and $ikG''u_z$, in Eq.~(\ref{eq:lin_u}). Being resistivity independent, these
terms compete directly with the classical inner-layer balance; in the induction
equation the relevant comparison is between $ik(Fu_z-Gb_z)+\xi u_z'$ and
$S^{-1}(b_z''-k^2b_z)$. Unlike the imposed-shear and externally
thinned backgrounds discussed in the Introduction, here the same $B_n$ generates
the linked $F$ and $G$ profiles and supplies the explicit $\xi$ terms in the
perturbation operator, so these ideal terms are part of the broadening mechanism
itself and can replace, rather than only time-modulate, the classical resistive
balance. The Somov--Verneta stabilization criterion $\xi\gtrsim S^{-3/4}$
\cite{1989SoPh..120...93S} was derived for a prescribed current layer from the
direct normal-field contribution alone, without the self-consistent
flow-advection and flow-curvature terms: in that fixed-equilibrium
ordering, normal-field crossing of the resonant layer outpaces resistive
diffusion across the classical tearing layer, so the resistive inner-layer
treatment is no longer the controlling approximation. In the evolving sheet, the
spatially varying flow terms provide additional ideal contributions
and can further reduce or replace the region in which resistivity dominates.

\subsection{Far-field spatial structure}

The far outer equations show explicitly that finite $\xi$ changes the structure
of the ideal solution. Taking $\mathrm{Pr}_m=0$ and neglecting the resistive
term, the far outer region satisfies $F\rightarrow1$, $F''\rightarrow0$,
$G\rightarrow0$, and $G''\rightarrow0$. Eliminating $u_z$ then gives
\begin{equation}
    \left[
        \left(\xi\frac{d}{dz}+ik\right)^2-\sigma^2
    \right]
    \left(
        \frac{d^2}{dz^2}-k^2
    \right)b_z = 0 .
    \label{eq:outer}
\end{equation}
Seeking far-field branches of the form $b_z\propto e^{\lambda z}$, where
$\lambda$ is the spatial eigenvalue, gives, for $\xi\neq0$,
\begin{equation}
\begin{aligned}
    b_z
    &=c_1e^{kz}+c_2e^{-kz} \\
    &\quad
    +c_3\exp\left[\left(-\frac{ik}{\xi}+\frac{\sigma}{\xi}\right)z\right]
    +c_4\exp\left[\left(-\frac{ik}{\xi}-\frac{\sigma}{\xi}\right)z\right].
\end{aligned}
\end{equation}
Assuming $k>0$, $\gamma>0$, and $\xi>0$, boundedness as $z\rightarrow+\infty$
removes the growing branches by setting $c_1=c_3=0$, so the physical far-field
solution contains both the standard tearing decay and a decaying normal-field
spatial component,
\begin{equation}
    b_z =
    c_2e^{-kz}
    +c_4\exp\left[-\frac{\sigma}{\xi}z-\frac{ik}{\xi}z\right],
    \qquad z>0 ,
    \label{eq:outer_solution}
\end{equation}
with the normal-field root chosen more generally so that
$\operatorname{Re}\lambda<0$. The finite-$\xi$ outer solution therefore does not
replace the classical tearing spatial solution; it supplements it
with an additional decaying oscillatory component, $\exp(-\gamma
z/\xi)\exp[-i(k+\omega)z/\xi]$ for $\sigma=\gamma+i\omega$. Whereas the usual
tearing component is localized on the scale $1/k$, this one has
envelope length $\ell_{\gamma,\xi}\sim\xi/\gamma$ and oscillatory length
$\ell_\xi\sim\xi/|k+\omega|$, corresponding to wavelength $2\pi\ell_\xi$. The
limit $\xi\rightarrow0$ is thus singular: the extra component
collapses into a rapidly oscillating boundary-layer contribution near the
matching region, while for finite $\xi$ it can remain spatially resolved.
These are two far-field spatial roots at a given temporal eigenvalue
$\sigma$; their coexistence is not, by itself, evidence for two distinct
temporal eigenbranches.

\subsection{Inner balance and transition scales}

Near the resonant surface the mathematical character of the problem
also changes. For $\xi=0$ and $G=0$, the ideal outer equation reduces to the
familiar singular form $b_z''-(k^2+F''/F)b_z=0$, whose solution can have a
finite jump in $b_z'$ across the resonant surface and is characterized by the
tearing parameter $\Delta'$. When $\xi\neq0$, the normal-field term introduces
higher spatial derivatives, schematically $\xi(b_z'''-k^2b_z')$, whose
coefficient does not vanish at $F=0$, while the terms proportional to $G$ and
$G''$ advect and distort the perturbation through the neutral layer. The
eigenfunction therefore need not be matched through a discontinuity in $b_z'$,
and the classical $\Delta'$-controlled tearing construction loses its strict
meaning once the resistive layer is removed.

The inner region is modified even more directly. Near the center, on
writing $s=w/a$,
{
\begin{equation}
\begin{aligned}
    F(z)&=F_1z+\mathcal{O}(z^3),\\
    G(z)&=G_0+\tfrac{1}{2}G_2z^2+\mathcal{O}(z^4),\\
    F_1&=a^{-1}\operatorname{sech}^2s,
    \quad G_0=\tanh s,\\
    G_2&=-2a^{-2}\tanh s\operatorname{sech}^2s .
\end{aligned}
    \label{eq:inner_profiles}
\end{equation}
}
The expansion has no term linear in $z$, since $G'(0)=0$. Defining
$H\equiv G-G_0$ and the comoving eigenvalue
$\widetilde{\sigma}\equiv\sigma+ikG_0$, with
$\widetilde{\omega}=\omega+kG_0$, transforms Eqs.~(\ref{eq:lin_u}) and
(\ref{eq:lin_b}) into
{
\begin{align}
    \widetilde{\sigma}\mathcal{D}u_z
    &=ik\left(F\mathcal{D}b_z-F''b_z\right)
    -ik\left(H\mathcal{D}u_z-G''u_z\right) \notag\\
    &\quad+\xi(\mathcal{D}b_z)'
    +\mathrm{Pr}_mS^{-1}\mathcal{D}^2u_z,
    \label{eq:comoving_u}\\
    \widetilde{\sigma}b_z
    &=ik\left(Fu_z-Hb_z\right)+\xi u_z'
    +S^{-1}\mathcal{D}b_z.
    \label{eq:comoving_b}
\end{align}
}
The corresponding leading inner induction equation is therefore
{
\begin{equation}
    \widetilde{\sigma}b_z
    \simeq ik\left(F_1zu_z-\frac{1}{2}G_2z^2b_z\right)
    +\xi u_z'+S^{-1}b_z'' .
    \label{eq:inner}
\end{equation}
}
Equation~(\ref{eq:inner}) retains both the reconnecting-field term and
the first nonuniform-flow correction; spatial variation of the flow enters
through $H=\mathcal{O}(z^2)$ and $G''$, with additional velocity-gradient
effects on the neutral-layer flanks.

To determine which additional term first invalidates the classical
resistive layer, we evaluate the new terms on the zeroth-order classical
solution. With $\widehat{\delta}_\eta\equiv\delta_\eta/a$ and $\gamma_{\rm cl}$
the layer width and growth rate of the classical branch under consideration, and
$\alpha\equiv ka$, the classical constant-$\psi$ FKR and nonconstant-$\psi$
Coppi scalings are
{
\begin{align}
    \widehat{\delta}_{\rm FKR}
    &\sim S^{-2/5}\alpha^{-2/5}(\Delta'a)^{1/5},\notag\\
    \gamma_{\rm FKR}\tau_A
    &\sim S^{-3/5}\alpha^{2/5}(\Delta'a)^{4/5},
    \label{eq:fkr_scales}\\
    \widehat{\delta}_{\rm Coppi}
    &\sim S^{-1/3}\alpha^{-1/3},\notag\\
    \gamma_{\rm Coppi}\tau_A
    &\sim S^{-1/3}\alpha^{2/3},
    \label{eq:coppi_scales}
\end{align}
}
respectively \cite{1963PhFl....6..459F,1976SvJPP...2..533C}. Here
$\Delta'$ is the matching parameter of the unperturbed classical outer problem;
for the initial Harris reference, $\Delta'_{\rm H}a=2(\alpha^{-1}-\alpha)$. We
use these scalings only as a perturbative onset test: they are self-consistent
while the ratios of the new terms to the classical terms are small, and equality
estimates where that ordering is lost, but they are not assumed to describe the
ideal-dominated layer beyond the transition. At finite $w$, the evolved profiles
and the explicit $\xi$ terms can change or remove the classical matching
structure, so Sec.~\ref{sec:representative_spectra} evaluates the actual balance
from the computed eigenfunctions.

We first consider the explicit normal-field term. Because $\xi$ is the
normal Alfv\'en speed normalized by $V_A$, its crossing rate through a classical
layer is $\xi/\widehat{\delta}_\eta$ in units of $\tau_A^{-1}$, and the
classical layer requires
$\xi/[(\gamma_{\rm cl}\tau_A)\widehat{\delta}_\eta]\ll1$; equality gives the
branch-dependent threshold
{
\begin{align}
    \xi_{\rm crit}(k)
    &\sim(\gamma_{\rm cl}\tau_A)\widehat{\delta}_\eta,
    \label{eq:xi_critical_general}\\
    \xi_{\rm crit}^{\rm FKR}
    &\sim S^{-1}\Delta'a,
    &\xi_{\rm crit}^{\rm Coppi}
    &\sim S^{-2/3}\alpha^{1/3}.
    \label{eq:xi_critical_branches}
\end{align}
}
The FKR and Coppi expressions meet at the fastest-growing scale and
both give $\xi_{\rm crit}\sim S^{-3/4}$, recovering the Somov--Verneta
ordering. This is the loss of the classical resistive-layer balance, not
necessarily the stabilization of the complete temporal spectrum.

We next ask whether the local spatial variation of $G$ can invalidate
the classical balance before this direct normal-field transition. With
$\mathcal{A}(s)\equiv\tanh s\operatorname{sech}^2s$, at $|z|\sim\delta_\eta$,
{
\begin{align}
    H&\sim-\mathcal{A}(s)\widehat{\delta}_\eta^2,
    &a^2G''(0)&=-2\mathcal{A}(s),
    \label{eq:inner_flow_scale}\\
    |T_H|&\sim
    \alpha\mathcal{A}(s)\widehat{\delta}_\eta^2|b_z|,
    &|T_{\rm res}|&\sim
    S^{-1}\widehat{\delta}_\eta^{-2}|b_z| .
    \label{eq:inner_flow_terms}
\end{align}
}
where $T_H$ is the local induction term proportional to $H=G-G_0$,
distinct from the finite velocity contrast sampled by the global mode below. Its
ratio to the resistive term is
{
\begin{equation}
    \frac{|T_H|}{|T_{\rm res}|}
    \sim S\alpha\mathcal{A}(s)\widehat{\delta}_\eta^4
    \sim
    \mathcal{A}(s)
    \begin{cases}
        \gamma_{\rm FKR}\tau_A/\alpha, & \text{FKR},\\
        \gamma_{\rm Coppi}\tau_A/\alpha, & \text{Coppi}.
    \end{cases}
    \label{eq:inner_flow_ratio}
\end{equation}
}
For $s\geq0$, $0\leq\mathcal{A}(s)\leq2/(3\sqrt{3})$, while the
slow-tearing ordering requires $\gamma_{\rm cl}\ll kV_A$, or
$\gamma_{\rm cl}\tau_A/\alpha\ll1$; at the classical fastest-growing scale,
where $\alpha\sim S^{-1/4}$, $\gamma\tau_A\sim S^{-1/2}$, and
$\widehat{\delta}_\eta\sim S^{-1/4}$, the ratio is $\mathcal{A}(s)S^{-1/4}$.
The locally quadratic flow variation is therefore asymptotically subdominant
wherever the classical reference solution remains self-consistent and does not
initiate the loss of the classical resistive balance. This conclusion does not
assume that the same ordering persists after the direct $\xi$ transition.

A complementary width estimate follows from differential advection
across the global mode. A tearing eigenfunction samples the central-flow region
and the flanks where $G$ falls toward zero, so the frame-invariant velocity
contrast across its global support is $\Delta G_{\rm mode}\sim G_0=\tanh(w/a)$,
and differential advection begins to distort the mode within one classical
e-folding time when $\alpha\Delta G_{\rm mode}\sim\gamma_{\rm cl}\tau_A$. This
gives
{
\begin{align}
    \tanh\left(\frac{w_{\rm adv}^{\rm FKR}}{a}\right)
    &\sim S^{-3/5}\alpha^{-3/5}(\Delta'a)^{4/5},
    \label{eq:wadv_fkr}\\
    \tanh\left(\frac{w_{\rm adv}^{\rm Coppi}}{a}\right)
    &\sim S^{-1/3}\alpha^{-1/3}.
    \label{eq:wadv_coppi}
\end{align}
}
Near the fastest-growing scale, either estimate reduces to
$w_{\rm adv}/a\sim S^{-1/4}$ when $w_{\rm adv}\ll a$. This is an onset scale for
global differential-advection effects, not a local replacement of resistivity
and not a stabilization width.

Section~\ref{subsec:tearing_structure} therefore establishes two
analytically distinct transitions: direct normal-field transport removes the
classical resistive-layer ordering near $\xi_{\rm crit}$, while differential
advection begins to distort the global mode near $w_{\rm adv}$. Neither
determines the identity or free-energy source of a growing temporal eigenmode.
We next evaluate the inner-balance transition in the computed spectrum and
treat the physical classification of the growing mode separately in
Sec.~\ref{sec:mode_classification}.

\section{Representative Spectra and the Inner-Balance Transition}
\label{sec:representative_spectra}

\subsection{Central-width diagnostic and spectral evolution}

\begin{figure*}
    \includegraphics[width=0.98\textwidth]{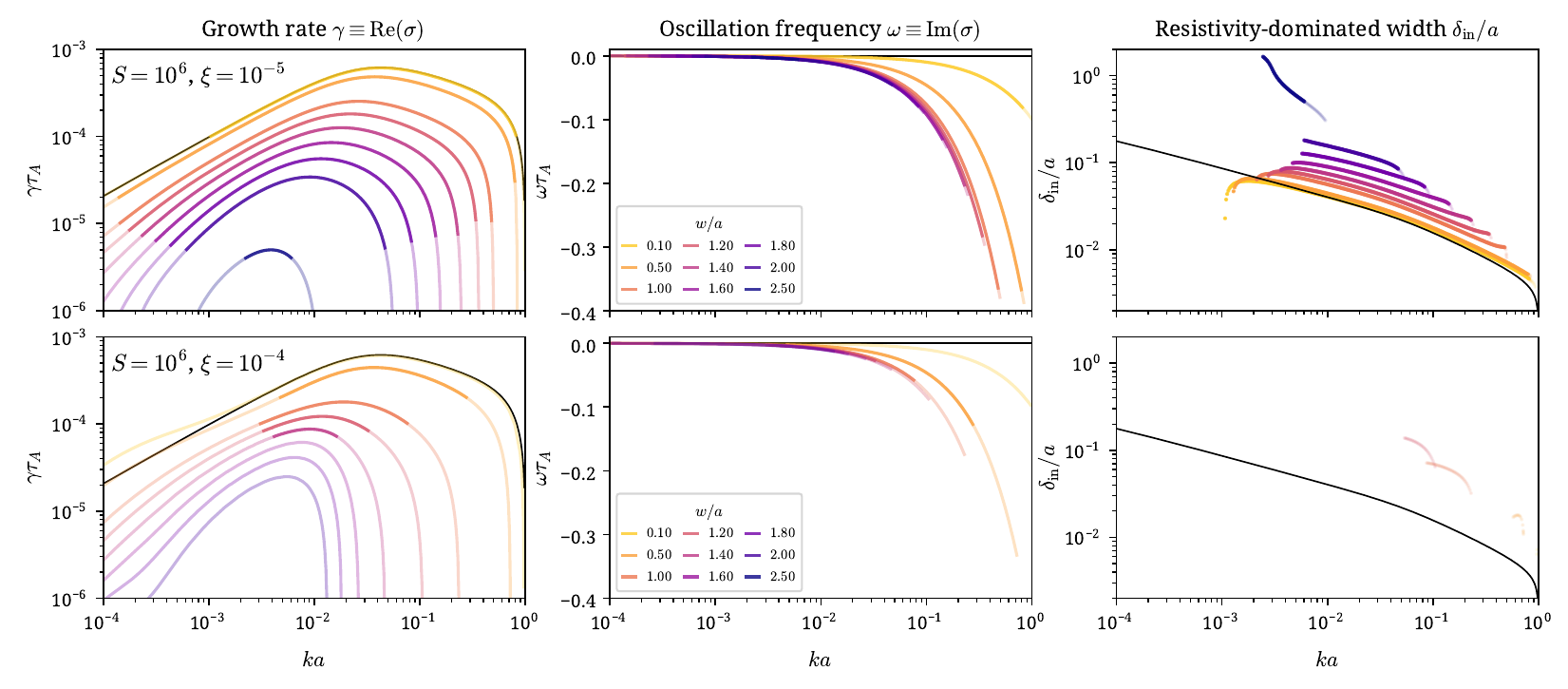}
    \caption{\label{fig:dispersions}
    Representative instantaneous spectra for $S=10^6$ and
    $\mathrm{Pr}_m=0$. The upper and lower rows correspond to
    $\xi=10^{-5}$ and $10^{-4}$, respectively, bracketing the prescribed-sheet
    estimate $\xi_{\rm crit}\sim S^{-3/4}$. Black curves denote the classical
    $(\xi,w/a)=(0,0)$ reference. From left to right, the panels show the
    growth rate $\gamma=\operatorname{Re}\sigma$, the oscillation frequency
    $\omega=\operatorname{Im}\sigma$, and the effective
    resistivity-dominated width $\delta_{\rm in}/a$, defined from the connected
    central region described below Eq.~(\ref{eq:complete_balance}). All
    panels use $w/a=0.10$, $0.50$, $1.00$, $1.20$, $1.40$, $1.60$, $1.80$,
    $2.00$, and $2.50$ and retain only modes satisfying $\gamma\tau_A S\geq1$,
    equivalent here to $\gamma\tau_A\geq10^{-6}$; faint portions show every
    mode so retained, and opaque portions those that also satisfy the local
    survival condition $\gamma\geq\xi/w$. The right column displays
    $\delta_{\rm in}$ only for modes with a connected central
    resistivity-dominated region. For a given
    $w/a$, no colored curve appears in any column if every mode fails the
    frozen-background criterion.}
\end{figure*}

We solve Eqs.~(\ref{eq:lin_u})--(\ref{eq:lin_b}) with \textsc{Psecas}
\cite{2019MNRAS.485..908B} on an infinite rational-Chebyshev grid based on
Chebyshev-Gauss nodes. The calculations use the \textsc{tearing-eigenmodes}
Python package (\url{https://github.com/gkowal/tearing-eigenmodes}) developed
by the first author, which implements the linearized equations and provides
high-level functions for calculating tearing eigenmodes using \textsc{Psecas}
as the eigensolver. For each mode, the rational-map scale and domain are
chosen so that at least 25 collocation points resolve the profile distance $w+a$
and the effective domain reaches the $1\%$ tail of the decaying outer solution;
the resolution is then increased from $N=256$ to $2048$ until the growth rate
satisfies absolute and relative tolerances of $10^{-10}$ and $10^{-5}$. We
retain modes satisfying the resistive frozen-background condition
$\gamma\tau_A S\geq1$, so that their growth time does not exceed the resistive
diffusion time of the background; because $w=\xi V_A t$, a broadened sheet with
$\xi=0$ does not belong to the normal-field-driven sequence. For $S=10^6$, the
prescribed-sheet estimate is $\xi_{\rm crit}\sim S^{-3/4}\simeq3\times10^{-5}$,
so the two finite normal fields $\xi=10^{-5}$ and $10^{-4}$ sample values about
a factor of three below and above it. This comparison uses $\xi_{\rm crit}$ only
to organize the cases; the evolved profiles need not have the same numerical
threshold.

We evaluate the induction balance in the frame moving with the
central flow to determine whether a central resistivity-dominated region
exists and, when it does, measure its half-width. Let $G_0=G(0)$. Removing this
constant advection speed changes the eigenvalue to
$\widetilde{\sigma}=\sigma+ikG_0$ and gives
{
\begin{align}
    T_F&=ikFu_z,
    &T_G^{\rm cf}&=-ik(G-G_0)b_z,\notag\\
    T_\xi&=\xi u_z',
    &T_{\rm res}&=S^{-1}(b_z''-k^2b_z),\notag\\
    T_{\rm ideal}^{\rm cf}&=T_F+T_G^{\rm cf}+T_\xi,
    &\widetilde{\sigma}b_z&=T_{\rm ideal}^{\rm cf}+T_{\rm res}.
    \label{eq:diagnostic_terms}
\end{align}
}
The ideal contribution is a signed complex sum: its terms are added
before the magnitude is taken. We define the local difference and its value at
the resonant surface,
{
\begin{equation}
\begin{aligned}
    T_d^{\rm cf}(z)&=|T_{\rm res}(z)|-|T_{\rm ideal}^{\rm cf}(z)|,\\
    \Phi_\eta(k,w)&=T_d^{\rm cf}(0).
\end{aligned}
    \label{eq:complete_balance}
\end{equation}
}
A central resistivity-dominated region exists only when
$\Phi_\eta>0$. Starting at $z=0$, we follow the connected interval over which
$T_d^{\rm cf}>0$; its first zero for $z>0$ defines the effective half-thickness
$\delta_{\rm in}$. If $\Phi_\eta\leq0$, we assign no central width, even if
numerical data contain an isolated resistive pocket farther from the resonant
surface. In a broadened profile the connected interval can include part of the
neutral region, so $\delta_{\rm in}$ is an effective resistivity-dominated width
of the eigenmode rather than necessarily a thin asymptotic inner layer. Along a
continued mode branch, a crossing $\Phi_\eta=0$ defines a mode-specific loss
width $w_\eta(k)$ where one exists. This crossing need not coincide with
$\gamma=0$. The perturbation-energy sources are examined separately in
Sec.~\ref{sec:mode_classification}.

Both rows of Fig.~\ref{fig:dispersions} show the same spectral
evolution, with the black $(\xi,w/a)=(0,0)$ curve included only as the classical
FKR/Coppi reference. As $w/a$ increases, the peak growth rate decreases and
moves toward smaller $ka$, and the laboratory-frame frequency becomes more
negative; Sec.~\ref{sec:mode_classification} quantifies the Doppler shift
and residual comoving frequency. Opaque segments identify
modes satisfying $\gamma\geq\xi/w$, and these bands contract as either $w/a$ or
$\xi$ increases: for $\xi=10^{-4}$ they remain visible only through
approximately $w/a\lesssim1.4$ and cluster around the instantaneous spectral
maximum, while the faint branches at larger widths remain unstable eigenmodes of
the frozen profile that fail the local survival condition.

The effective resistivity-dominated width changes strongly with $w/a$.
For $\xi=10^{-5}$, a connected central region survives over a finite wavenumber
band at every displayed width, and along the fastest-growing sequence
$\delta_{\rm in}/a$ rises from $0.0245$ at $w/a=0.1$ to $0.161$ at $w/a=2$ and
$0.768$ at $w/a=2.5$; the last value is comparable to the sheet scale and
therefore describes an effective resistivity-dominated width, not a thin
asymptotic inner layer. For $\xi=10^{-4}$, the fastest-growing mode has
$\Phi_\eta\leq0$ at every displayed width with an eligible spectrum: at
$w/a=0.1$ only two sampled modes near $ka\simeq0.97$ have $\Phi_\eta>0$,
restricted higher-$k$ bands with connected central regions also occur at
$w/a=0.5$, $1.0$, and $1.2$, no central region occurs for $1.4\leq w/a\leq2.0$,
and at $w/a=2.5$ every mode falls below the growth-rate cutoff. All modes with a
finite width in this row lie on faint curve segments and fail the local survival
condition. The sign of $\Phi_\eta$ is therefore mode and width dependent: a
finite width can reappear for a broadened mode without restoring the classical
Harris-sheet ordering, and growing modes can remain in the instantaneous
spectrum after that ordering has been lost. Their physical classification is
treated separately in Sec.~\ref{sec:mode_classification}.

\subsection{Laboratory-frame induction balance}

\begin{figure*}
    \includegraphics[width=\textwidth]{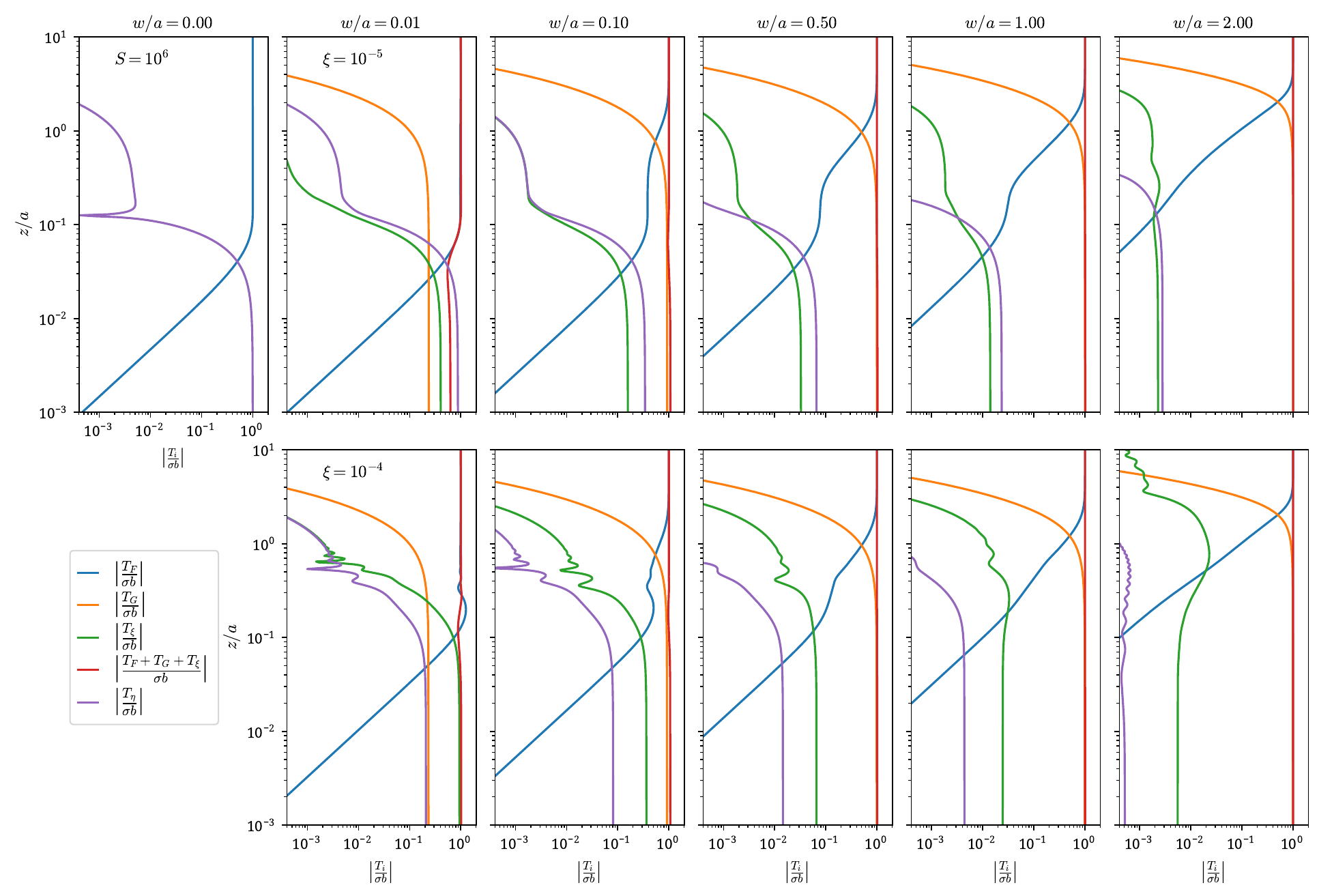}
    \caption{\label{fig:term_balance}
    Separate laboratory-frame induction contributions for
    $S=10^6$, $\mathrm{Pr}_m=0$, and $ka=0.01$.
    The plot in the first column is the common classical Harris reference
    $(\xi,w/a)=(0,0)$. The upper and lower rows use $\xi=10^{-5}$ and $10^{-4}$,
    respectively, through $w/a=0.01$, $0.10$, $0.50$, $1.00$, and $2.00$. Curves
    show $|T_F/(\sigma b_z)|$, $|T_G/(\sigma b_z)|$, $|T_\xi/(\sigma b_z)|$,
    $|(T_F+T_G+T_\xi)/(\sigma b_z)|$, and $|T_\eta/(\sigma b_z)|$, with
    the terms defined in Eq.~(\ref{eq:explicit_induction_terms}). The red curve
    is the magnitude of the signed complex ideal sum, not an envelope of the
    individual magnitudes. The plotted $T_G=-ikG(z)b_z$ contains the full
    laboratory-frame advection; the subtraction of $G_0$ in
    Eq.~(\ref{eq:diagnostic_terms}) is used only to measure the central
    resistivity-dominated width.}
\end{figure*}

To compare the individual induction terms, we retain the
laboratory frame used by the eigenvalue problem:
{
\begin{align}
    T_F&=ikFu_z,
    &T_G&=-ikGb_z,\notag\\
    T_\xi&=\xi u_z',
    &T_\eta&=S^{-1}(b_z''-k^2b_z),\notag\\
    \sigma b_z&=T_F+T_G+T_\xi+T_\eta.
    \label{eq:explicit_induction_terms}
\end{align}
}
Here $T_\eta=T_{\rm res}$, whereas $T_G$ contains the full
laboratory-frame advection. As in Eq.~(\ref{eq:diagnostic_terms}), the net
ideal contribution is the signed complex sum $T_F+T_G+T_\xi$, whose magnitude
is taken only after the terms have been added; the individual magnitudes are
therefore not additive fractions of $|\sigma b_z|$.

The classical first column of Fig.~\ref{fig:term_balance} recovers the
familiar division of labor: $T_\eta$ controls the immediate vicinity of the
resonant surface, while $T_F$ takes over farther out. The $w/a=0.01$ column
isolates the direct normal-field effect before substantial broadening: at
$\xi=10^{-5}$, $T_\eta$ remains the largest individual contribution at the
resonant surface, whereas at $\xi=10^{-4}$, $T_\xi$ is already larger than
$T_\eta$. This reversal establishes the initial loss of central resistive
dominance in the higher-$\xi$ sequence and is consistent with its absence along
the fastest-growing modes in Fig.~\ref{fig:dispersions}; it does not require
$\Phi_\eta$ to have the same sign for every $k$ as the profiles broaden.

The remaining columns show the additional effect of broadening. As
$w/a$ increases, the normalized resistive contribution decreases, the full
laboratory-frame term $T_G=-ikG(z)b_z$ approaches order unity across the central
region, and $T_F$ becomes dominant again outside it. The term $T_G$ contains
the central-flow Doppler contribution $-ikG_0b_z$. Its order-unity normalized
magnitude does not establish whether the velocity gradient supplies perturbation
energy. We subtract $G_0$ only for the central-width diagnostic in
Eq.~(\ref{eq:diagnostic_terms}); Sec.~\ref{sec:mode_classification} tests the
energy source independently.

Figure~\ref{fig:term_balance} thus attributes the initial loss of
central resistive dominance to $T_\xi$, whereas at larger widths the evolution
of $\Phi_\eta$ and the displacement of the spectral peak depend on the complete
broadened $F$ and $G$ profiles and on the changing eigenfunction. A positive
effective width that reappears for one broadened mode does not restore the
classical FKR or Coppi ordering, and a mode that remains unstable after the
central resistivity-dominated region has disappeared need not obey the classical
resistive tearing balance. Broadening modifies both profiles: the central
reconnecting-field gradient $F'(0)=a^{-1}\operatorname{sech}^2(w/a)$ decreases
as $w/a$ grows, while the central-flow speed $G(0)=\tanh(w/a)$ increases and
velocity gradients develop on the neutral-layer flanks. Over the sampled widths,
the peak growth rate decreases and the fastest-growing wavenumber moves toward
smaller $ka$.

These calculations test the mechanism for representative cases; they
do not constitute a parameter survey. Near the classical fastest-growing scale,
the estimate for the direct normal-field transition is $\xi S^{3/4}=O(1)$, so
increasing $S$ lowers the estimated threshold in $\xi$. The differential-advection
estimates in Eqs.~(\ref{eq:wadv_fkr}) and (\ref{eq:wadv_coppi}) give
$w_{\rm adv}/a\sim S^{-1/4}$ near the same scale when $w_{\rm adv}\ll a$.
The onset of differential advection and the mode-specific loss width
$w_\eta(k)$, defined by a crossing $\Phi_\eta(k,w)=0$ where one exists, are
based on different criteria and need not coincide. Neither criterion alone
establishes stabilization of the full instantaneous spectrum. At fixed $w/a$,
increasing $\xi$ strengthens the explicit ideal normal-field terms in
Eqs.~(\ref{eq:lin_u}) and (\ref{eq:lin_b}) but also changes
the time at which the sheet reaches that width, since $w=\xi V_A t$, so a
fixed-$w/a$ comparison in $\xi$ is not the time history of one sheet. The
initial half-thickness $a$ enters the normalized eigenproblem through
$S=aV_A/\eta$, $w/a$, and $ka$. At fixed $V_A$, $\eta$, $\xi$, $t$, and dimensional
$k$, increasing $a$ raises $S$, reduces $w/a$, and raises $ka$. Finite
$\mathrm{Pr}_m$ enters the frozen-profile eigenproblem through the viscous
coefficient $\mathrm{Pr}_mS^{-1}$
and is expected to change the classical viscoresistive inner-layer scalings and
the quantitative transition values, but not the ideal origin of the
self-broadening solution; because Fig.~\ref{fig:dispersions} uses
$\mathrm{Pr}_m=0$, we make no quantitative claim about viscosity here.

\section{Physical Character of the Growing Mode Beyond the
Resistive-Layer Regime}
\label{sec:mode_classification}

The crossing $\Phi_\eta=0$ changes the local induction balance; it does
not, by itself, mark a bifurcation at which tearing becomes a different temporal
instability. Classifying the growing solution requires three complementary
tests: its structural connection to classical tearing, its intrinsic frequency
after removal of the central-flow Doppler shift, and the source of its
perturbation energy. Each admits an analytical constraint before the
eigenfunctions are examined.

\subsection{Analytical classification criteria}

The first criterion concerns the structural connection to classical
tearing. $F$ is odd and $G$ is even, so Eqs.~(\ref{eq:lin_u}) and
(\ref{eq:lin_b}) preserve two parity subspaces, and the classical tearing
subspace, with even $b_z$ and odd $u_z$, remains an exact symmetry class for
every $\xi$ and $w$. Numerical parity residuals test whether a computed
eigenfunction lies in this class, but parity alone cannot establish that it is
the continuation of the classical tearing eigenmode; that requires comparison of
the eigenfunctions along the parameter sequence.

Secondly, the transformation in Eqs.~(\ref{eq:comoving_u}) and
(\ref{eq:comoving_b}) is exact and separates the Doppler contribution $-kG_0$
from the intrinsic frequency $\widetilde{\omega}=\omega+kG_0$. A nonzero
laboratory-frame $\omega$ therefore does not by itself identify an Alfv\'enic
instability, and, as shown in Sec.~\ref{subsec:tearing_structure},
Eq.~(\ref{eq:outer_solution}) is a spatial root at a specified $\sigma$ rather
than an additional temporal eigenbranch.

Finally, the perturbation-energy budget identifies the terms that can
support exponential growth. Multiplying the linearized momentum and induction
equations by the conjugate perturbations, integrating over $z$, and imposing decay
at infinity gives the exact two-dimensional budget
{
\begin{equation}
    \begin{split}
        2\gamma E&=P_F+P_G+P_\xi-D_\eta-D_\nu,\\
        E&=\frac{1}{2}\int
        (|\boldsymbol{u}|^2+|\boldsymbol{b}|^2)\,dz .
    \end{split}
    \label{eq:energy_budget}
\end{equation}
}
where
{
\begin{equation}
\begin{aligned}
    P_F&=\int F'\operatorname{Re}
    (u_x^*b_z-b_x^*u_z)\,dz,\\
    P_G&=\int G'\operatorname{Re}
    (b_x^*b_z-u_x^*u_z)\,dz,\\
    P_\xi&=\xi\int\operatorname{Re}\left(
    \boldsymbol{u}^*\cdot\boldsymbol{b}'
    +\boldsymbol{b}^*\cdot\boldsymbol{u}'\right)\,dz \notag\\
    &=\xi\left[\operatorname{Re}
    (\boldsymbol{u}^*\cdot\boldsymbol{b})
    \right]_{-\infty}^{+\infty}=0,\\
    D_\eta&=S^{-1}\int
    |\boldsymbol{\nabla}\times\boldsymbol{b}|^2\,dz,\\
    D_\nu&=\mathrm{Pr}_m S^{-1}\int
    |\boldsymbol{\nabla}\times\boldsymbol{u}|^2\,dz .
\end{aligned}
\label{eq:energy_terms}
\end{equation}
}
The two terms in $P_\xi$ are the normal-field contributions to the
kinetic- and magnetic-energy equations; their sum corresponds to the boundary flux
shown in Eq.~(\ref{eq:energy_terms}), which vanishes for decaying eigenfunctions.
The normal field can therefore exchange kinetic and magnetic perturbation energy
and alter the eigenfunction and inner balance, but supplies no net perturbation
energy, and retaining $P_\xi$ in the numerical budget tests this cancellation
directly.

\begin{figure*}
    \includegraphics[width=0.8\textwidth]{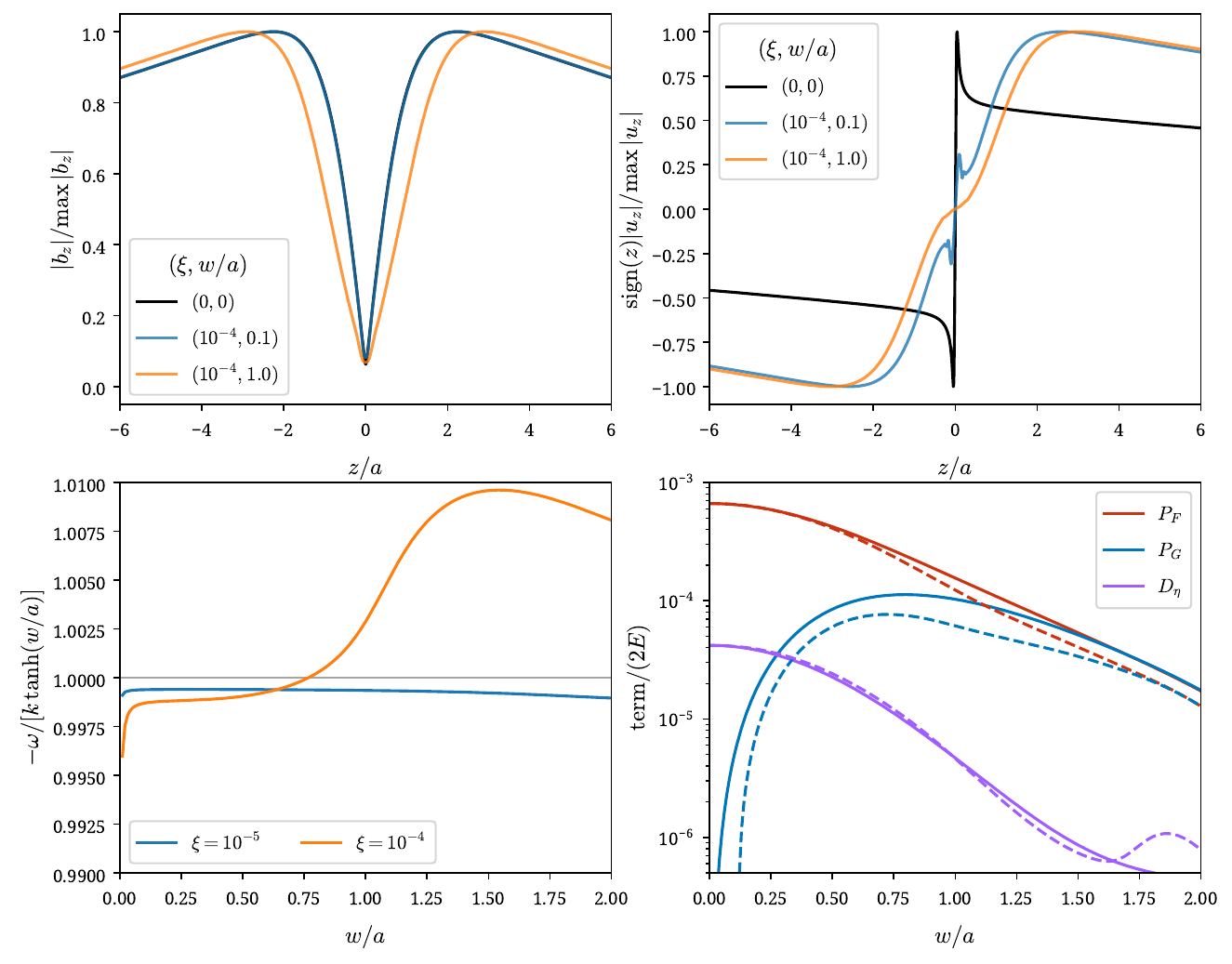}
    \caption{\label{fig:branch_diagnostics}
    Diagnostics of the structure, intrinsic frequency, and
    free-energy source of the growing mode for $S=10^6$ and
    $\mathrm{Pr}_m=0$. The upper-left and upper-right panels show the normalized
    magnetic and signed velocity amplitudes, respectively, at fixed
    $ka=0.04266$ for the classical reference and for
    $(\xi,w/a)=(10^{-4},0.1)$ and $(10^{-4},1)$. The bottom-left panel shows the
    laboratory frequency divided by the Doppler frequency of the central flow,
    $-\omega/(kG_0)$, where $G_0=\tanh(w/a)$; values near unity correspond to a
    small comoving frequency $\widetilde{\omega}=\omega+kG_0$. The bottom-right
    panel shows $P_F$, $P_G$, and the resistive dissipation $D_\eta$, normalized
    by $2E$ and plotted on a logarithmic scale. These quantities are the
    individual contributions to the growth-rate balance in
    Eq.~(\ref{eq:energy_budget}). The bottom panels follow the fastest-growing
    mode at each displayed $w/a$.
    Blue and orange curves in the bottom-left panel denote $\xi=10^{-5}$ and
    $10^{-4}$, respectively; solid and dashed curves in the bottom-right panel
    denote $\xi=10^{-5}$ and $10^{-4}$, respectively, while color identifies
    the energy term. The logarithmic ordinate displays only positive values.
    The unplotted conservative contribution $P_\xi$ remains near zero, and the
    relative energy-budget residual is below $1.5\times10^{-5}$.}
\end{figure*}

Both $D_\eta$ and $D_\nu$ are nonnegative dissipation rates; the
calculations below use $\mathrm{Pr}_m=0$, for which $D_\nu=0$, but the general
budget makes the role of finite viscosity explicit. Only $P_F$ and $P_G$ can
supply net perturbation energy, coupling the perturbations to the
reconnecting-field and equilibrium-velocity gradients, respectively, with
positive values representing extraction from the background and negative values
a return of energy to it. Their relative magnitudes depend on the eigenfunctions,
so the numerical classification requires three quantities not fixed by the
analytical constraints: structural overlap with classical tearing, the residual
comoving frequency, and the partition between $P_F$ and $P_G$, which can be
evaluated numerically as follows.

\subsection{Numerical evaluation}

We first quantify structural continuity. For a perturbation variable
$q$, define the phase-independent overlap $\mathcal{O}_q=|\int q_0^*q\,dz|/
(\int|q_0|^2dz\int|q|^2dz)^{1/2}$, where $q_0$ is the $\xi=0$, $w=0$ reference at
the same $ka=0.04266$. For $\xi=10^{-4}$ and $w/a=(0.01,0.1,1)$, the magnetic
overlaps are $\mathcal{O}_{b_z}=(1.000,1.000,0.996)$ and the velocity overlaps
are $\mathcal{O}_{u_z}=(0.984,0.963,0.953)$; as required by the exact symmetry,
the complex eigenfunctions have even $b_z$ and odd $u_z$ to relative residuals
below $3\times10^{-10}$, which serves as a numerical consistency check. The large
overlaps and the profiles in the upper panels of
Fig.~\ref{fig:branch_diagnostics} show that the sampled growing solution remains
structurally close to classical tearing after the inner balance ceases to be
resistivity dominated.

The comoving-frequency diagnostic separates intrinsic propagation from
advection. For the fastest modes in both rows of Fig.~\ref{fig:dispersions} and
$0.01\leq w/a\leq2$, the bottom-left panel of Fig.~\ref{fig:branch_diagnostics}
gives $-\omega/[k\tanh(w/a)]=0.996$--$1.010$, so the laboratory-frame propagation
is predominantly the Doppler shift of the central flow while the intrinsic
comoving frequency is much smaller.

The energy budget then identifies the source of growth. Dividing
Eq.~(\ref{eq:energy_budget}) by $2E$ expresses $\gamma$ as the signed sum of the
plotted contributions. The bottom-right panel of
Fig.~\ref{fig:branch_diagnostics} shows that the reconnecting-field contribution
$P_F/(2E)$ is dominant at early stages of the sheet evolution, i.e. at small and
moderate $w/a$. At later stages, i.e. at larger $w/a$, the contribution of the
velocity shear term $P_G/(2E)$ becomes comparable. Over this larger-width portion
of the sequence, both production terms, as well as the resistive dissipation
$D_\eta/(2E)$, decrease as the sheet broadens. The conservative contribution
$P_\xi/(2E)$ is not plotted because it remains numerically negligible. Along
the same fastest-mode sequences, $\gamma\tau_A$ decreases monotonically from
$6.2\times10^{-4}$ at $w/a=0$ to $3.4\times10^{-5}$ and $2.5\times10^{-5}$ at
$w/a=2$ for $\xi=10^{-5}$ and $10^{-4}$, respectively. Thus, although the
velocity gradient supplies an increasing share of the positive energy input,
the total energy production per unit perturbation energy continues to fall.

The two fastest-mode sequences underlying the bottom panels of
Fig.~\ref{fig:branch_diagnostics} are each tabulated at 201 equally spaced widths
over $0\leq w/a\leq2$, and a composite Simpson integration of the corresponding
$\gamma_{\max}$ values gives $N_{\max}(2)=58.7$ for $\xi=10^{-5}$ and $5.12$ for
$\xi=10^{-4}$. Because $k_{\max}$ changes with $w/a$, these values bound the
corresponding fixed-$k$ instantaneous-eigenvalue integrals rather than the exact
amplification of a time-dependent perturbation. For $\xi=10^{-4}$, the
fastest-mode growth time $t_\gamma=\gamma_{\max}^{-1}$ first exceeds the
characteristic flow-development time $t_{\rm sat}\simeq\tau_A/\xi$ at
$w/a\simeq1.32$, where $N_{\max}=4.73$, and the sequence then leaves the local
survival regime at $w/a\simeq1.59$, where $t_\gamma=t_w$ and $N_{\max}=4.95$;
only $0.175$ additional upper-envelope e-foldings accumulate by $w/a=2$. In
contrast, the $\xi=10^{-5}$ sequence satisfies $t_\gamma<t_{\rm sat}$ and
$t_\gamma<t_w$ throughout $0.1\leq w/a\leq2$. At $w/a=2$,
$\gamma_{\max}\tau_A S=34.3$ and $24.9$ for $\xi=10^{-5}$ and $10^{-4}$,
respectively, so the growth time in neither sequence reaches the resistive
diffusion time $t_\eta=S\tau_A$ in the calculated interval.

\section{Discussion and Conclusions}

Our results refine the usual statement that a normal field stabilizes
tearing. In a prescribed sheet, stabilization follows from direct normal-field
crossing of the resistive layer \cite{1989SoPh..120...93S}; in the
evolving sheet, the same field also changes the background on which tearing
develops. The reconnecting profile broadens and a central sheet-parallel flow
develops with oppositely signed velocity gradients on two flanks, which for the
profiles studied here lowers the maximum instantaneous growth rate and shifts
the fastest-growing mode toward smaller $ka$, hence toward larger longitudinal
scales. The ideal normal-field and velocity-profile terms can then overtake
resistivity in the inner induction balance: for $\xi=10^{-4}$, the
fastest-growing mode has no connected central resistivity-dominated region at
any eligible displayed width.

Where these ideal terms dominate, they remove the classical resistive
ordering without establishing a distinct ideal temporal branch. In the sampled
fixed-$k$ sequence, the tearing parity and large eigenfunction overlaps instead
support a normal-field-modified continuation of classical tearing; the
fastest-growing modes have a predominantly Doppler-shifted laboratory-frame
frequency, and their small comoving frequency does not support a distinct
Alfv\'enic interpretation. The energy budget gives
$P_\xi\simeq0$, as required for the conservative normal-field contribution, with
extraction from the reconnecting-field gradient dominant at small and moderate
widths and the velocity-gradient contribution comparable only at broad widths,
after the growth rate has decreased. We therefore use ``ideal'' only for the
dominant local inner terms; establishing an ideal temporal instability would
require continuing the same branch in $S$ and finding a nonzero limiting growth
rate as $S\rightarrow\infty$, which is beyond the present calculation.

For $(S,\xi)=(10^6,10^{-4})$, although the normal field is only
$B_n=10^{-4}B_0$, the fastest-mode growth time first exceeds $t_{\rm sat}$ at
$w/a\simeq1.32$. When the neutral-layer half-width reaches only $w/a\simeq1.59$,
$\gamma_{\max}$ is about ten times smaller than at $w/a=0$ and $t_\gamma=t_w$, so
further broadening outpaces modal growth; by $w/a=2$, $\gamma_{\max}$ has
decreased by a factor of about 25 and only $0.175$ additional upper-envelope
e-foldings have accumulated after the crossing. Further linear amplification is
therefore nearly arrested over the sampled interval, even though the
frozen-profile spectrum remains unstable and the nonlinear outcome remains seed
dependent. For $\xi=10^{-5}$, $t_\gamma$ remains shorter than both $t_{\rm sat}$
and $t_w$ throughout $0.1\leq w/a\leq2$, and neither sequence reaches the
separate resistive frozen-background limit $t_\gamma=t_\eta$.

The spectra displayed here are intentionally selective and serve as
representative tests of this mechanism rather than as a parameter survey; a
complementary numerical manuscript is devoted to systematic exploration of $S$,
$\xi$, $w/a$, $a$, and $\mathrm{Pr}_m$ together with nonlinear calculations, and
the present manuscript does not use those undisplayed results as evidence. For
a local, approximately planar MHD region with sufficiently uniform and
persistent $B_n$, fixed-Harris disruption estimates may therefore miss
a background-evolution channel that changes the instantaneous stability
problem as the sheet evolves.

The present calculation is restricted to the linear stability of
frozen profiles. It determines instantaneous eigenvalues and eigenfunctions,
branch structure, and whether the induction balance contains a
resistivity-dominated region; it does not evolve perturbation amplitudes or
determine nonlinear saturation, island or plasmoid formation, the reconnection
rate, or the final topology. In classical tearing the end of the linear stage is
commonly estimated by $W_{\rm isl}\sim\delta_{\rm in}$, when the island
half-width becomes comparable to the resistive inner-layer width
\cite{2016PhPl...23j0702C}; we do not evaluate this seed-dependent condition,
which would require following each fixed-$k$ branch through the evolving
background. Disappearance of $\delta_{\rm in}$ in Fig.~\ref{fig:dispersions},
whose induction contributions are compared in Fig.~\ref{fig:term_balance},
therefore means that this classical criterion ceases to apply, not that islands
or all nonlinear structures are excluded. Wave propagation and nonlinear
interactions may also modify an inhomogeneous background and generate small
scales: Meringolo et al. demonstrated the joint action of phase mixing and
nonlinear cascade for Alfv\'enic perturbations in three-dimensional
compressible-MHD simulations of a coronal-loop model \cite{Meringolo2024}. Their
geometry differs from the present current sheet, so we use that result only as
context for nonlinear processes omitted here; determining how such processes
affect island growth and the final topology requires a time-dependent nonlinear
MHD calculation, reserved for the complementary numerical study.

Our results do not question the existence of plasmoid-mediated reconnection in
the regimes where its ordering assumptions are satisfied. Rather, they show that
the applicability of those assumptions is itself a dynamical question:
the background can leave the classical resistive-layer ordering as it
evolves, and its evolution can outpace the weakening mode before $\gamma$
vanishes, while the subsequent nonlinear outcome remains outside the present
analysis.

The principal conclusions are: (i) a normal field modifies tearing
both directly and indirectly through the self-consistent broadening and
sheet-parallel flow of the background; (ii) the
decrease of the maximum growth rate principally reflects the weakening of
perturbation-energy production from the reconnecting-field gradient as the
neutral layer broadens (Fig.~\ref{fig:branch_diagnostics}). The flank-shear
contribution becomes positive and partly offsets this loss, but it does not
reverse the monotonic decrease of $\gamma_{\max}$ with $w/a$, while the
fastest-growing mode shifts toward larger longitudinal scales; (iii) dominance
of the ideal inner terms removes the classical resistive-layer ordering, while
the sampled growing modes remain a tearing-like continuation rather than a
demonstrated new ideal temporal branch; and (iv) for $S=10^6$ and a normal field
of only $B_n=10^{-4}B_0$,
$\gamma_{\max}$ is about ten times smaller than at $w/a=0$ once the neutral-layer
half-width reaches $w/a\simeq1.59$, after which less than $0.2$ additional
upper-envelope e-folds accumulate by $w/a=2$. The frozen-profile growth rate
remains positive, and this conclusion about further linear amplification does not
exclude finite-resistivity island or plasmoid formation.

\begin{acknowledgments}
G.K. and D.F.-G. acknowledge support from FAPESP grants 2013/10559-5,
2021/02120-0, 2021/06502-4, and 2022/03972-2. The numerical calculations were
performed using the computational resources of the Group of Theoretical
Astrophysics at EACH-USP.

We acknowledge the use of large-language-model tools in the preparation
of this work. OpenAI GPT-5.5, GPT-5.6, and GPT-6 were used to automate the
analysis of the numerical results against criteria defined by the authors,
such as verifying that all computed growth rates exceed the rate of resistive
expansion of the current sheet and identifying modes that had not reached the
required tolerance so that they could be recomputed at higher resolution.
GPT-5.5 assisted the initial derivation of the far-field solution, which the
authors subsequently verified by hand, and the linearized equations were
verified using SymPy. Anthropic Claude Opus 5.5 was used to refactor and polish
the final version of the tearing-eigenmodes Python package; the code was
reviewed and tested by the authors. All interpretation of the results and the
conclusions drawn from them are the authors' own.
\end{acknowledgments}

\section*{Data Availability}
The numerical eigenmode data supporting this study are available from the
authors upon reasonable request. The \textsc{Psecas} source code
(\url{https://github.com/gkowal/psecas}) and the modifications used for these
calculations (\url{https://github.com/gkowal/tearing-eigenmodes}) are publicly
available.

\bibliography{ms}

\end{document}